# Elementary quantum gates in different bases

Sergey A. Podoshvedov

*Department of General and Theoretical Physics, South Ural State University, Lenin Av. 76, Chelyabinsk, Russia*
e-mail: sapo66@mail.ru

We introduce transformation matrix connecting sets of the displaced states with different displacement amplitudes. Arbitrary pure one-mode state can be represented in new basis of the displaced number (Fock) states ($\alpha-$ representation) by multiplying the transposed transformation matrix on a column vector of initial state. Analytical expressions of the $\alpha-$ representation of superposition of vacuum and single photon and two-mode squeezed vacuum (TMSV) are obtained. On the basis of the developed mathematical formalism, we consider the mechanism of interaction between qubits which is based on their displaced properties. Superposed coherent states deterministically displace target state on equal modulo but opposite on sign values. Registration of the single photon in auxiliary mode (probabilistic operation) results in constructive interference and gives birth to entangled hybrid state corresponding to outcome of elementary quantum gates. The method requires minimal number of resource and works in realistic scenario.

.
## 1. Introduction

Quantum information with qubits as carriers provides a totally new way of information processing to execute intriguing tasks which would not be possible by means of traditional classical methods [1]. Quantum parallelism and entanglement are fundamental features of quantum mechanics which provide a significant speed-up for certain computational tasks such as factorizing a large integer [2] and unsorted database searching [3]. A deterministic Bell-state measurement cannot be performed using linear optics and photon detection [4]. Only two of the four Bell states can be identified with photon qubits and success probability of the measurement does not exceed half. Possible increase of the success probability of the Bell measurement [5] requires a heavy increase of resources. Measurement-based linear optical quantum operators look attractive with theoretical point of view [6]. But practical realization of the cluster states is hardly possible. A proposal in [7] needs medium only with nonlinearity responsible for cross-phase modulation that maybe not feasible. So, optical fiber has cubic nonlinearity that contributes to comparable self- and cross modulation of the propagating optical fields.

A superposition of coherent states (SCS) with the same amplitude but opposite phases is an example of another logical qubit. One says about even and odd superposition of coherent states (SCSs) in dependency on phase relation between components of the superposition. The SCSs are orthogonal, can be recognized as optical analogue of the cat states [8] and can be exploited for quantum information processing [9]. But it is hardly possible to control the coherent qubits in deterministic way as nonlinear effects in existing materials are extremely small and not enough. Beam splitter with high transmittance provides an alternative probabilistic mechanism (subtraction of photons) to achieve effective nonlinear effect on input state when reflected mode is measured. Such measurement-based probabilistic technique is realized in many practical proposals [10-13]. Probabilistic realization of a qubit rotation and a nontrivial controlled-NOT gate is considered in [14, 15].

It is natural to consider other possible optical states and to develop new ways of working with them for the successful processing of quantum information. The technique of the qubit control can be extended [16-18] by means of use of displaced number states of the quantum oscillator [19]. Displaced states are characterized by additional classic parameter being displacement amplitude which can be called by their size. If the size of the displaced state is large enough, then it may be even considered as macroscopic. The displaced states become number ones with the exact number of photons in the case of zeroth displacement amplitude. As the number states, and their displaced versions can be represented by the same error circles on the phase plane with quadrature components as axes. The difference between the number state and its displaced counterpart on the phase plane is that the centers of the circles are shifted relative to each other by the amount of the difference of their sizes. But sizes of their images on the phase plane are equal to each other. Displacement of the state on the phase plane provides the displacement operator that is deterministically implemented in quantum optics and which at least does not reduce the probability of the final event. The displaced number states may contain infinite number of elementary particles distributed over some law. Nevertheless, we can also work with the displaced number state as with one particle having size and neglect its internal structure [20, 21].

In this paper, mathematical apparatus to work with displaced states is developed. The basis for this method is the transformation matrix which connects the sets of the displaced states of different sizes. Matrix elements are obtained. The transformation matrix can be the basis for the introduction of the new state representation in terms of the displaced number states. Given the representation, new mechanism of the qubit interaction is considered. The mechanism means something that makes the input qubits to become entangled and this something depends on the properties of qubits. We note only that the initial qubits are significantly different in size. One of the qubits deterministically performs the displacement operation. The qubit can be considered as a macroscopic. Deterministic operation of displacement is changed by measurement in auxiliary mode. The output state becomes entangled and the entanglement can be called a hybrid because of the difference in sizes of the participating qubits. We show that the output states correspond to the output states of the one- and two-qubit elementary gates. We note only that distinct initial qubits are used on input into the gates. The gates are probabilistic but require a minimum amount of resources for their implementation.

## 2. $\alpha-$ representation of pure states

Quantization of the electromagnetic field is accomplished by choosing classical complex amplitudes to be mutually adjoint operators. The dynamical behavior of the electric-field amplitudes is described by an ensemble of independent harmonic oscillators obeying well-known communication relations. Respectively, the number (Fock) states of the harmonic oscillator are used for description. The number states are the partial case of the displaced number ones with zeroth displacement amplitude. The number and displaced number states are connected with each other by unitary transformation. Therefore, any pure state can be also represented in terms of the displaced number states. We are going to call the decomposition over displaced number states $\alpha-$ representation of the state. $\alpha-$ representation can be also introduced for the mixed states but it is out of the consideration.

### 2.1 Transformation matrix for displacement operator

The eigenvalues of Hamiltonian of the quantum harmonic oscillator are defined by integers $(n=0,1,2,...,\infty)$. Corresponding eigenstates $|n\rangle$ are known as number or Fock states [22]. The number states are orthogonal $\langle n|m\rangle = \delta_{nm}$, where $\delta_{nm}=1$, if $n=m$, otherwise $\delta_{nm} = 0$. Displaced number states are obtained by additional application of the displacement operator [22]

$$D(\alpha) = \exp(\alpha a^+ - \alpha^* a), \tag{1}$$

to the number state $|n\rangle$ [17, 18]

$$|n,\alpha\rangle = D(\alpha)|n\rangle, \tag{2}$$

where $\alpha$ is an amplitude of the displacement and $a$, $a^+$ are the bosonic annihilation and creation operators. Set of the displaced number states is complete for arbitrary value of $\alpha$. We can also introduce annihilation and creation operators $A = a - \alpha$ and $A^+ = a^+ - \alpha^*$ for the displaced number states by analogy with bosonic operators. Their application to the displaced number stare $|n,\alpha\rangle$ yields

$$A|n,\alpha\rangle = \sqrt{n}|n-1,\alpha\rangle, \tag{3}$$

$$A^+|n,\alpha\rangle = \sqrt{n+1}|n+1,\alpha\rangle. \tag{4}$$

Then, the displaced Fock states can be also defined as eigenstates of operator $A^+A$

$$A^+A|n,\alpha\rangle = n|n,\alpha\rangle. \tag{5}$$

The mean energy of the displaced number state in non-dimensional units (neglecting vacuum fluctuations)

$$\langle n,\alpha|a^+a|n,\alpha\rangle = n + |\alpha|^2, \tag{6}$$

becomes the sum of the number state (quantum) energy $n$ and classical wave intensity $|\alpha|^2$. The displaced number states (2) are defined by two numbers: quantum discrete number $n$ and classical continuous parameter $\alpha$ which can be recognized as their size. Somewhat, it may testify about manifestation of particle-wave duality of the states, especially, of the coherent state or the same displaced vacuum. Although, the number and displaced number states have some quantum noise properties in common, they are not physically similar to each other as shown in Fig. 1.

Choose two sets of orthogonal displaced number states

$$\{|n,\alpha\rangle, n = 0,1,2,...,\infty\}, \tag{7}$$

$$\{|n,\alpha'\rangle, n = 0,1,2,...,\infty\}, \tag{8}$$

where, in general case, $\alpha \neq \alpha'$. Every element from one set can be expressed through states from another for its completeness. Coefficients of the decomposition are the inner products [23].

$$\langle n,\alpha'|n,\alpha\rangle \tag{9}$$

which takes as input two vectors $|n,\alpha\rangle$ and $|n,\alpha'\rangle$ from sets (7,8) and produces, in general case, a complex number as output [1]. Let us present another method of derivation of the coefficients. There exists the following chain of transformations with operators and states

$$|l,\alpha'\rangle = D(\alpha')|l\rangle = \exp\left(-|\alpha'|^2/2\right)\exp(\alpha'a^+)\exp(-\alpha'^*a)|l\rangle =$$

$$\exp\left(-|\alpha'|^2/2\right)\exp(\beta a^+)\exp(\alpha a^+)\exp(-\alpha^*a)\exp(\alpha^*a)\exp(-\alpha'a)|l\rangle =$$

$$\exp\left(-|\alpha'|^2/2\right)\exp(\beta a^+)\exp(\alpha a^+)\exp(-\alpha^*a)\exp((\alpha-\alpha')^*a)|l\rangle =$$

$$\exp\left(-\left(|\alpha'|^2-|\alpha|^2\right)/2\right)\exp(\beta a^+)\exp(-|\alpha|^2/2)\exp(\alpha a^+)\exp(-\alpha^*a)\exp((\alpha-\alpha')^*a)|l\rangle$$

$$= \exp\left(-\left(|\alpha'|^2-|\alpha|^2\right)/2\right)\exp(\beta a^+)D(\alpha)\exp((\alpha-\alpha')^*a)|l\rangle =$$

$$\exp\left(-\left(|\alpha'|^2-|\alpha|^2\right)/2\right)\exp(\beta a^+)D(\alpha)$$
$$\left(1+(\alpha-\alpha')^*a+(\alpha-\alpha')^{*2}a^2/2!+(\alpha-\alpha')^{*3}a^3/3!+\ldots+(\alpha-\alpha')^{*l}a^l/l!\right)|l\rangle =$$

$$\exp\left(-\left(|\alpha'|^2-|\alpha|^2\right)/2\right)\exp(\beta a^+)D(\alpha)$$
$$\begin{pmatrix}|l\rangle+(\alpha-\alpha')^*\sqrt{l}|l-1\rangle+(\alpha-\alpha')^{*2}\sqrt{l(l-1)}|l-2\rangle/2!+\\(\alpha-\alpha')^{*3}\sqrt{l(l-1)(l-2)}|l-3\rangle/3!+\ldots+(\alpha-\alpha')^{*l}\sqrt{l!}|0\rangle/l!\end{pmatrix} =$$

$$\exp\left(-\left(|\alpha'|^2-|\alpha|^2\right)/2\right)\exp(\beta a^+)$$
$$\begin{pmatrix}|l,\alpha\rangle+(\alpha-\alpha')^*\sqrt{l}|l-1,\alpha\rangle+(\alpha-\alpha')^{*2}\sqrt{l(l-1)}|l-2,\alpha.\rangle/2!+\\(\alpha-\alpha')^{*3}\sqrt{l(l-1)(l-2)}|l-3,\alpha\rangle/3!+\ldots+(\alpha-\alpha')^{*l}\sqrt{l!}|0,\alpha\rangle/l!\end{pmatrix}$$ , (10)

where $\alpha' = \alpha + \beta$ ($\beta = \alpha' - \alpha$) and $\alpha$, $\beta$ are the arbitrary numbers and $*$ is an operation of complex conjugation. Expression (10) consists of $l+1$ terms. It is possible to show the following identity

$$\exp(\beta a^+)|l,\alpha\rangle = \sum_{n=0}^{\infty}\frac{(\beta a^+)^n}{n!}|l,\alpha\rangle =$$
$$\exp(\beta\alpha^*)\sum_{n=0}^{\infty}\frac{\beta^n}{\sqrt{n!}}\sqrt{\frac{(n+1)(n+2)\ldots(n+l)}{l!}}|n+l,\alpha\rangle$$ (11)

holds. Briefly mention how to derive the formula (11). First consider the effect of each member $(\beta a^+)^n/n!$ on the state $|l,\alpha\rangle$. It produces a superposition of $n+1$ displaced states $|l+m,\alpha\rangle$, where quantity $m$ is changed in the range from 0 to $n$. The next step is to collect all the terms which belong to the displaced state $|l+m,\alpha\rangle$. The collected terms turns out to form infinite series being phase factor $\exp(\beta\alpha^*)$. The common factor can be imposed for sum that gives the final expression (11). Inserting expression (11) into (10), one obtains final representation of the displaced number state $|l,\alpha'\rangle$ in terms of the states $|l,\alpha\rangle$.

$$|l, \alpha'\rangle = \exp\left(-\left(|\alpha'|^2 + |\alpha|^2\right)/2\right)\exp(\alpha'\alpha^*)$$

$$\begin{pmatrix} \sum_{n=0}^{\infty} \frac{(\alpha' - \alpha)^n}{\sqrt{n!}} \sqrt{\frac{(n+1)(n+2)...(n+l)}{l!}} |n+l, \alpha\rangle + \\ (\alpha - \alpha')^* \sqrt{l} \sum_{n=0}^{\infty} \frac{(\alpha' - \alpha)^n}{\sqrt{n!}} \sqrt{\frac{(n+1)(n+2)...(n+l-1)}{(l-1)!}} |n+l-1, \alpha\rangle + \\ (\alpha - \alpha')^{*2} \sqrt{l(l-1)}/2! \sum_{n=0}^{\infty} \frac{(\alpha' - \alpha)^n}{\sqrt{n!}} \sqrt{\frac{(n+1)(n+2)...(n+l-2)}{(l-2)!}} |n+l-2, \alpha\rangle + \\ (\alpha - \alpha')^{*3} \sqrt{l(l-1)(l-2)}/3! \sum_{n=0}^{\infty} \frac{(\alpha' - \alpha)^n}{\sqrt{n!}} \sqrt{\frac{(n+1)(n+2)...(n+l-3)}{(l-3)!}} |n+l-3, \alpha\rangle + \\ .... + \\ (\alpha - \alpha')^{*(l-1)} \sqrt{l(l-1)(l-2)...2}/(l-1)! \sum_{n=0}^{\infty} \frac{(\alpha' - \alpha)^n}{\sqrt{n!}} \sqrt{n+1} |n+1, \alpha\rangle + \\ (\alpha - \alpha')^{*(l)} \sqrt{l!}/l! \sum_{n=0}^{\infty} \frac{(\alpha' - \alpha)^n}{\sqrt{n!}} |n, \alpha\rangle \end{pmatrix}. \quad (12)$$

The decomposition (12) enables to introduce transformation matrix $U$ connecting input and output base states

$$\begin{vmatrix} |0, \alpha'\rangle \\ |1, \alpha'\rangle \\ |2, \alpha'\rangle \\ |3, \alpha'\rangle \\ |4, \alpha'\rangle \\ |5, \alpha'\rangle \\ |6, \alpha'\rangle \\ ... \\ |n, \alpha'\rangle \\ ... \end{vmatrix} = U \begin{vmatrix} |0, \alpha\rangle \\ |1, \alpha\rangle \\ |2, \alpha\rangle \\ |3, \alpha\rangle \\ |4, \alpha\rangle \\ |5, \alpha\rangle \\ |6, \alpha\rangle \\ ... \\ |n, \alpha\rangle \\ ... \end{vmatrix} = F \begin{vmatrix} c_{00} & c_{01} & c_{02} & c_{03} & c_{04} & c_{05} & c_{06} & ... & c_{0m} & ... \\ c_{10} & c_{11} & c_{12} & c_{13} & c_{14} & c_{15} & c_{16} & ... & c_{1m} & ... \\ c_{20} & c_{21} & c_{22} & c_{23} & c_{24} & c_{25} & c_{26} & ... & c_{2m} & ... \\ c_{30} & c_{31} & c_{32} & c_{33} & c_{34} & c_{35} & c_{36} & ... & c_{3m} & ... \\ c_{40} & c_{41} & c_{42} & c_{43} & c_{44} & c_{45} & c_{46} & ... & c_{4m} & ... \\ c_{50} & c_{51} & c_{52} & c_{53} & c_{54} & c_{55} & c_{56} & ... & c_{5m} & ... \\ c_{60} & c_{61} & c_{62} & c_{63} & c_{64} & c_{65} & c_{66} & ... & c_{6m} & ... \\ ... & ... & ... & ... & ... & ... & ... & ... & ... & ... \\ c_{n0} & c_{n1} & c_{n2} & c_{n3} & c_{n4} & c_{n5} & c_{n6} & ... & c_{nm} & ... \\ ... & ... & ... & ... & ... & ... & ... & ... & ... & ... \end{vmatrix} \begin{vmatrix} |0, \alpha\rangle \\ |1, \alpha\rangle \\ |2, \alpha\rangle \\ |3, \alpha\rangle \\ |4, \alpha\rangle \\ |5, \alpha\rangle \\ |6, \alpha\rangle \\ ... \\ |n, \alpha\rangle \\ ... \end{vmatrix}, \quad (13)$$

where $F = \exp\left(-\left(|\alpha|^2 + |\alpha'|^2\right)/2\right)\exp(\alpha'\alpha^*)$, matrix elements $c_{kl} = c_{kl}(\alpha', \alpha)$ depend on input and output values $\alpha'$, $\alpha$ and follow from (12). Matrix $U$ has its reverse $U^{-1}$ since the displacement operator (1) is unitary $U^{-1} = U^+$, where $U^+$ is a linear matrix known as Hermitian conjugate of the matrix $U$. If we take the value of $\alpha = 0$, we obtain the expansion of displaced states on the number states. It is possible consider the opposite case of the expansion of the number states on their displaced counterparts even without referring to the inverse matrix conversion in the case of $\alpha' = 0$.

Consider the case of $\alpha' = 0$. The decomposition

$$|l\rangle = |\Psi_1\rangle + |\Psi_2\rangle \quad (14)$$

follows from (12,13), where a wave function $|\Psi_1\rangle$ is written as

$$|\Psi_1\rangle = \exp(-|\alpha|^2/2)\sum_{n=0}^{l-1} c_{ln}|n,\alpha\rangle, \qquad (15)$$

for the values $0 \le n < l$. Wave amplitudes of the superposition (15) stems from (12)

$$c_{ln} = \frac{(\alpha^*)^{l-n}}{\sqrt{l!}\sqrt{n!}}\sum_{k=0}^{n}(-1)^k C_n^k |\alpha|^{2k} \prod_{k}^{n-1}(l-n+k+1), \qquad (16)$$

with $C_n^k = n!/(k!(n-k)!)$ being binomial coefficients and symbol $\Pi$ means product of the numbers. Wave function $|\Psi_2\rangle$ is an infinite superposition of the displaced number states for $l \le n < \infty$

$$|\Psi_2\rangle = \exp(-|\alpha|^2/2)\sum_{n=m}^{\infty} c_{ln}|n,\alpha\rangle, \qquad (17)$$

with the wave amplitudes

$$c_{ln} = \frac{(-1)^{n-l}\alpha^{n-l}}{\sqrt{l!}\sqrt{n!}}\sum_{k=0}^{l}(-1)^k C_l^k |\alpha|^{2k} \prod_{k}^{l-1}(n-l+k+1). \qquad (18)$$

Consider big values of the displacement amplitude $\alpha \gg 1$ to regard the states (2) with more energy as macroscopic. The decomposition $U^{-1}$ can be fully recognized natural with classical point of view since it reflects that macroscopic state (object) is made up of microscopic states (stuff). Probability to look for the number states with large value of $n$ in the superposition is negligible. But contrary idea that microscopic state (object) is composed of macroscopic states (with more energy) seems at least weird and downright bizarre with classical point of view. Nevertheless, it doesn't contradict quantum mechanics rules. Both direct and reverse transformations are legal in quantum mechanics.

Probability to observe $n$ photons displaced on $\alpha$ in $l$ – photon state is defined by the corresponding wave amplitude $c_{ln}$ modulo squared

$$P_{ln}(\alpha) = |c_{ln}(\alpha)|^2. \qquad (19)$$

Normalization condition $\langle l|l\rangle = 1$ for arbitrary $l$ – photon state

$$\sum_{n=0}^{\infty} P_{ln}(\alpha) = 1, \qquad (20)$$

is implemented. Figures 2(a-f) reveal probability distributions of single photon over the displaced number states for the different displacement amplitudes. The distributions manifest typical inherent oscillatory features especially with increase of the displacement amplitude, when $l+1$ peaks are beheld in $l$ – photon state.

## 2.2 $\alpha$ – representation of arbitrary state

Consider arbitrary state

$$|\Psi\rangle = \sum_{m=0}^{n} a_m |m,\alpha'\rangle =$$
$$\left\| |0,\alpha'\rangle \quad |1,\alpha'\rangle \quad |2,\alpha'\rangle \quad ... \quad |m,\alpha'\rangle \quad ... \right\| \| a_0 \quad a_1 \quad a_2 \quad ... \quad a_m \quad ... \|^T, \qquad (21)$$

where $a_k$ are the state's amplitudes and upper limit $n$ in sum (21) may take values from 0 up to $\infty$ and symbol $T$ means matrix transposition. Expression (21) is the product of the vector row of the basic states on the vector column of amplitudes. The state information can be also encoded in other basis as

$$|\Psi\rangle = \sum_{m=0}^{n} a_m |m,\alpha'\rangle = ||0,\alpha'\rangle \quad |1,\alpha'\rangle \quad |2,\alpha'\rangle \quad ... \quad |m,\alpha'\rangle \quad ...|| a_0 \quad a_1 \quad a_2 \quad ... \quad a_m \quad ...|^T =$$

$$\sum_{m=0}^{n} b_m |m,\alpha\rangle = ||0,\alpha\rangle \quad |1,\alpha\rangle \quad |2,\alpha\rangle \quad ... \quad |m,\alpha\rangle \quad ...|| b_0 \quad b_1 \quad b_2 \quad ... \quad b_m \quad ...|^T$$

(22)

Here, the column vectors at different bases are related with each other

$$\begin{vmatrix} b_0 \\ b_1 \\ b_2 \\ ... \\ b_m \\ ... \end{vmatrix} = FU^T \begin{vmatrix} a_0 \\ a_1 \\ a_2 \\ ... \\ a_m \\ ... \end{vmatrix} \tag{23}$$

through the transformation matrix $U$ (13) and $U^T$ is the transposed matrix. Thus, the same state (21) can be represented by either column vector

$$a = | a_0 \quad a_1 \quad a_2 \quad ... \quad a_m \quad ...|^T \tag{24}$$

in basis (8) ($\alpha'$ − representation ) or column vector

$$b = | b_0 \quad b_1 \quad _2 \quad ... \quad b_m \quad ...|^T \tag{25}$$

in basis (7) ($\alpha$ − representation).

The physical information is contained in wave amplitudes of the initial state. All what we wish to know about physical system in quantum mechanics is in column vector of the state amplitudes of the system (21). Formula (23) shows there are different ways (24,25) of encoding the same information enclosed in the state (21) in dependency on choice of the base states. It may resemble $x-$ and $p-$ representations of the wave function of the quantum particle. Note the decomposition of the state in terms of the Fock states can be called $0-$representation since $\alpha = 0$. All we need to make use of information in different bases is at least set of the projective operators on the displaced number states

$$M_n = |n,\alpha\rangle\langle n,\alpha|. \tag{26}$$

To physically realize the operation, displacement operator (1) with amplitude $-\alpha$ is applied to initial state before measurement as shown in figure 3(a). The displacement operation in optics can be effectively and deterministically performed by means of mixing of initial state with coherent state of large amplitude [9] on highly transmitting beam splitter (Fig. 3(b)). The following measurement of the state can be implemented by avalanche photodiodes (APD). APD possesses high quantum efficiency but can only discriminate the presence of radiation from the vacuum. It can be used to reconstruct the photon statistics but cannot be used as photon counters. Number state $|n\rangle$ can be measured by photon number resolving detector (PNRD) able to distinguish outcomes from different number states. Now, PNRD are only prototype with restricted possibilities [24].

### 2.3 $\alpha-$ representation of superposition of vacuum and single photon

Consider the following balanced superpositions of vacuum and single photon

$$|\Delta_\pm\rangle = (|0\rangle \pm |1\rangle)/\sqrt{2}. \tag{27}$$

The states (27) are orthogonal to each other. Column vectors of the states (in $0-$representation) are given by

$$a_{\pm} = (1/\sqrt{2})|1 \quad \pm 1 \quad 0 \quad ... \quad 0 \quad ...|^T. \tag{28}$$

Using the formula (23), one constructs output column vector (25) of the states in basis of the displaced number states ($\alpha$ – representation of the states)

$$|\Delta_+\rangle = \frac{\exp(-|\alpha|^2/2)}{\sqrt{2}}\left(\begin{array}{l}(1+\alpha^*)|0,\alpha\rangle + \\ \sum_{n=1}^{\infty}\frac{(-1)^n \alpha^{n-1}}{\sqrt{n!}}(\alpha - (n - |\alpha|^2))|n,\alpha\rangle\end{array}\right), \tag{29}$$

$$|\Delta_-\rangle = \frac{\exp(-|\alpha|^2/2)}{\sqrt{2}}\left(\begin{array}{l}(1-\alpha^*)|0,\alpha\rangle + \\ \sum_{n=1}^{\infty}\frac{(-1)^n \alpha^{n-1}}{\sqrt{n!}}(\alpha + (n - |\alpha|^2))|n,\alpha\rangle\end{array}\right). \tag{30}$$

Column vectors following from (29,30) are again converted to (28) in the case of $\alpha = 0$. The wave amplitudes of the states (27) in arbitrary $\alpha$ – representation satisfy the condition

$$b_{2n}^{(+)}(\alpha) = b_{2n}^{(-)}(-\alpha), \tag{31}$$

$$b_{2n+1}^{(+)}(\alpha) = -b_{2n+1}^{(-)}(-\alpha), \tag{32}$$

where superscripts $\pm$ regard positive $|\Delta_+\rangle$ and negative $|\Delta_-\rangle$ states, respectively. Probability distribution over $n$ – number states displaced on $\alpha$ is defined by

$$P_{n\pm}(\alpha) = \frac{\exp(-|\alpha|^2)|\alpha|^{2(n-1)}}{2 \quad n!}|\alpha \mp (n - |\alpha|^2)|^2, \tag{33}$$

which is even function with regard to sign change of the parameter $\alpha$

$$P_{n+}(\alpha) = P_{n-}(-\alpha). \tag{34}$$

Normalization condition leads to

$$\sum_{n=0}^{\infty} P_{n\pm}(\alpha) = 1. \tag{35}$$

## 2.4 $\alpha$ – representation of two-mode squeezed state

Let us derive $\alpha$ – representation of the TMSV defined through two-mode squeezed operator

$$|\Psi\rangle_{12} = S_{12}(r)|00\rangle_{12} = (1/\cosh r)\sum_{n=0}^{\infty}(\tanh r)^n |n\rangle_1|n\rangle_2 = (1/\cosh r)A^T B, \tag{36}$$

where operator $S_{12}(r)$ is given by [22]

$$S_{12}(r) = \exp(r(a_1^+ a_2^+ - a_1 a_2)) \tag{37}$$

with $r$ being the squeezing parameter and infinite vector columns are the following

$$A^T = |\;|0\rangle \quad \tanh r|1\rangle \quad (\tanh r)^2|2\rangle \quad ... \quad (\tanh r)^l|l\rangle \quad ...\;|, \tag{38}$$

$$B = |\;|0\rangle \quad |1\rangle \quad |2\rangle \quad ... \quad |n\rangle \quad ...\;|^T. \tag{39}$$

Present the vector column $B$ as product of the transformation matrix $U$ (13) on the vector column of the displaced number states. It enables to rewrite $A^T B$ as

$$A^T B = \exp(-|\alpha|^2/2) \left| |0\rangle \quad \tanh r |1\rangle \quad (\tanh r)^2 |2\rangle \quad ... \quad (\tanh r)^l |l\rangle \quad ... \right|$$

$$\begin{vmatrix} c_{00} & c_{01} & c_{02} & ... & c_{0l} & ... \\ c_{10} & c_{11} & c_{12} & ... & c_{1l} & ... \\ c_{20} & c_{21} & c_{22} & ... & c_{2l} & ... \\ ... & ... & ... & ... & ... & ... \\ c_{l0} & c_{l1} & c_{l2} & ... & c_{ll} & ... \\ ... & ... & ... & ... & ... & ... \end{vmatrix} \begin{Vmatrix} |0,\alpha\rangle \\ |1,\alpha\rangle \\ |2,\alpha\rangle \\ ... \\ |l,\alpha\rangle \\ ... \end{Vmatrix} . \qquad (40)$$

Introduce the following infinite vector row of the states

$$C^T = \left| |\psi_0\rangle \quad |\psi_1\rangle \quad |\psi_2\rangle \quad ... \quad |\psi_n\rangle \quad ... \right| =$$

$$\left| |0\rangle \quad \tanh r |1\rangle \quad (\tanh r)^2 |2\rangle \quad ... \quad (\tanh r)^l |l\rangle \quad ... \right| \begin{vmatrix} c_{00} & c_{01} & c_{02} & ... & c_{0l} & ... \\ c_{10} & c_{11} & c_{12} & ... & c_{1l} & ... \\ c_{20} & c_{21} & c_{22} & ... & c_{2l} & ... \\ ... & ... & ... & ... & ... & ... \\ c_{l0} & c_{l1} & c_{l2} & ... & c_{ll} & ... \\ ... & ... & ... & ... & ... & ... \end{vmatrix}, \qquad (41)$$

to finally present TMSV (36) in the form

$$|\Psi\rangle_{12} = \left(\exp(-|\alpha|^2/2)/\cosh r\right) C^T B = \left(\exp(-|\alpha|^2/2)/\cosh r\right) \sum_{n=0}^{\infty} |\psi_n\rangle_1 |n,\alpha\rangle_2 , \qquad (42)$$

where the wave function $|\psi_n\rangle$ is infinite superposition of the number states

$$|\psi_n\rangle = \sum_{l=0}^{\infty} c_{ln}(\alpha)(\tanh r)^l |l\rangle . \qquad (43)$$

Expression (42) may be called $\alpha$ – representation of the TMSV but it is hardly comfortable for practical applications. We are going to manipulate with the states (43) to obtain more convenient form of the $\alpha$ – representation of the TMSV.

The wave amplitudes of the superposition (43) are formed from elements of $n$ column of the transformation matrix (13) multiplied by factors $(\tanh r)^l$. Contribution of the additional factor occurs from squeezing operator (37). It is possible to demonstrate the wave amplitude $c_{ln}(\alpha)(\tanh r)^l$ for arbitrary value of the parameter $l$ can be decomposed into finite series of the basic functions

$$\{c^*_{l0}(\beta), c^*_{l1}(\beta), c^*_{l2}(\beta), c^*_{l3}(\beta), ..., c^*_{ln}(\beta)\}, \qquad (44)$$

where $\beta = \alpha^* \tanh r$. Set of the base functions (44) consists of $n+1$ terms. Mathematical proof of the fact is not trivial and is not presented here. The argument of the base functions (44) is a product of $\alpha^*$ by $\tanh r$, while amplitudes $c_{ln}(\alpha)$ are the functions of only $\alpha$. We have the following decomposition of $c_{ln}(\alpha)(\tanh r)^l$ on the base functions

$$c_{ln}(\alpha)(\tanh r)^l = \sum_{k=0}^{n} a_{nk} c^*_{lk}(\beta), \qquad (45)$$

where amplitudes of the decomposition $a_{ln}(\alpha)$ are calculated from the matrix equations that are not presented here for their complexity. The amplitudes are given by

$$a_{n0}(\alpha) = \alpha^n \left((\tanh r)^2 - 1\right)^n / \sqrt{n!}, \qquad (46)$$

$$a_{n1}(\alpha) = n\alpha^{n-1} \tanh r \left((\tanh r)^2 - 1\right)^{n-1} / \sqrt{n!}, \qquad (47)$$

$$a_{nk}(\alpha) = n(n-1)(n-2)...(n-k)\alpha^{n-k}(\tanh r)^k ((\tanh r)^2 - 1)^{n-k} / (\sqrt{k!}\sqrt{n!}), \qquad (48)$$

$$a_{nn}(\alpha) = (\tanh r)^n. \qquad (49)$$

Let us make use of the following symbolic matrix

$$\begin{vmatrix} & a_{n0} & a_{n1} & a_{n2} & a_{n3} & a_{n4} & ... & a_{nn} & |\psi_n\rangle \\ |0\rangle & c_{00}^*(\beta) & c_{01}^*(\beta) & c_{02}^*(\beta) & c_{03}^*(\beta) & c_{04}^*(\beta) & ... & c_{0n}^*(\beta) & c_{0n}(\alpha) \\ |1\rangle & c_{10}^*(\beta) & c_{11}^*(\beta) & c_{12}^*(\beta) & c_{13}^*(\beta) & c_{14}^*(\beta) & ... & c_{1n}^*(\beta) & c_{1n}(\alpha)\tanh r \\ |2\rangle & c_{20}^*(\beta) & c_{21}^*(\beta) & c_{22}^*(\beta) & c_{23}^*(\beta) & c_{24}^*(\beta) & ... & c_{2n}^*(\beta) & c_{2n}(\alpha)(\tanh r)^2 \\ |3\rangle & c_{30}^*(\beta) & c_{31}^*(\beta) & c_{32}^*(\beta) & c_{33}^*(\beta) & c_{34}^*(\beta) & ... & c_{3n}^*(\beta) & c_{3n}(\alpha)(\tanh r)^3 \\ |4\rangle & c_{40}^*(\beta) & c_{41}^*(\beta) & c_{42}^*(\beta) & c_{43}^*(\beta) & c_{44}^*(\beta) & ... & c_{4n}^*(\beta) & c_{4n}(\alpha)(\tanh r)^4 \\ |5\rangle & c_{50}^*(\beta) & c_{51}^*(\beta) & c_{52}^*(\beta) & c_{53}^*(\beta) & c_{54}^*(\beta) & ... & c_{5n}^*(\beta) & c_{5n}(\alpha)(\tanh r)^5 \\ ... & ... & ... & ... & ... & ... & ... & ... & ... \\ |k\rangle & c_{k0}^*(\beta) & c_{k1}^*(\beta) & c_{k2}^*(\beta) & c_{k3}^*(\beta) & c_{k4}^*(\beta) & ... & c_{kn}^*(\beta) & c_{kn}(\alpha)(\tanh r)^k \\ ... & ... & ... & ... & ... & ... & ... & ... & ... \\ |n\rangle & c_{n0}^*(\beta) & c_{n1}^*(\beta) & c_{n2}^*(\beta) & c_{n3}^*(\beta) & c_{n4}^*(\beta) & ... & c_{nm}^*(\beta) & c_{nn}(\alpha)(\tanh r)^n \\ ... & ... & ... & ... & ... & ... & ... & ... & ... \\ |l\rangle & c_{l0}^*(\beta) & c_{l1}^*(\beta) & c_{l2}^*(\beta) & c_{l3}^*(\beta) & c_{l4}^*(\beta) & ... & c_{\ln}^*(\beta) & c_{\ln}(\alpha)(\tanh r)^l \\ & |0,\beta\rangle & |1,\beta\rangle & |2,\beta\rangle & |3,\beta\rangle & |4,\beta\rangle & ... & |n,\beta\rangle & |\psi_n\rangle \end{vmatrix} \qquad (50)$$

to elucidate details of the next transformation. The matrix (50) consists of finite number $n+3$ of columns and infinite number of the rows. First column of the matrix (50) is a column of number states from 0 up to $\infty$. Last column is a vector column of the state $|\psi_n\rangle$ (43). First row involves amplitudes of the decomposition (46-49) over the base functions (44) and also the state (43). Next rows except for elements of the first and last columns involve the base functions (44). Comparing the column elements under corresponding amplitude $a_{nm}$ with $m$ varying from 0 up to $n$ and rows of the reverse transformation matrix (13) for $\alpha' = 0$, we can notice they are identical. Thus, the state (43) can be rewritten as finite $n+1$ superposition of the displaced number states

$$|\psi_n\rangle = \exp(|\beta|^2/2)a_{n0}\sum_{k=0}^{n}b_{nk}|k,\beta\rangle = \exp(|\beta|^2/2)a_{n0}D(\beta)\sum_{k=0}^{n}b_{nk}|k\rangle, \qquad (51)$$

where
$$b_{n0}(\alpha) = 1, \qquad (52)$$
$$b_{n1}(\alpha) = a_{n1}(\alpha)/a_{n0}(\alpha) = n\gamma, \qquad (53)$$
$$b_{nk}(\alpha) = a_{nk}(\alpha)/a_{n0}(\alpha) = n(n-1)(n-2)...(n-k+1)\gamma^k/\sqrt{k!}, \qquad (54)$$
$$b_{nn}(\alpha) = a_{nn}(\alpha)/a_{n0}(\alpha) = \sqrt{n!}\gamma^n, \qquad (55)$$
and
$$\gamma = \tanh r / (\alpha((\tanh r)^2 - 1)). \qquad (56)$$

Consider the finite sum of the number states

$$\sum_{k=0}^{n} b_{nk}|k\rangle. \qquad (57)$$

It is operator $(1 + \gamma a^+)$ in power $n$ acting on vacuum state

$$\sum_{k=0}^{n} b_{nk} |k\rangle = (1 + \gamma a^{+})^{n} |0\rangle. \tag{58}$$

The following identity

$$a_{n0} \gamma^{n} = (\tanh r)^{n} / \sqrt{n!} \tag{59}$$

takes place. Finally, rewrite the state $|\psi_n\rangle$ (43) as

$$|\psi_n\rangle = \exp(|\beta|^2 / 2) D(\beta) (\tanh r)^n (a^+ - \delta^*)^n |0\rangle / \sqrt{n!}, \tag{60}$$

where the parameter $\delta$ is determined by

$$\delta = \alpha^* (1 - (\tanh r)^2) / \tanh r. \tag{61}$$

Then, $\alpha$ – representation of the two-mode squeezed state (43) can be presented as

$$|\Psi\rangle_{12} = \left(\exp(-(\sinh r)^2 |\delta|^2 / 2)/\cosh r\right) D_1(\alpha^* \tanh r) D_2(\alpha)$$
$$\sum_{n=0}^{\infty} \left((\tanh r)^n (a_1^+ - \delta^*)^n |0\rangle_1 / \sqrt{n!}\right) |n\rangle_2 = \tag{62}$$
$$\left(\exp(-(\sinh r)^2 |\delta|^2 / 2)/\cosh r\right) D_1(\alpha^* \tanh r) D_2(\alpha) \sum_{n=0}^{\infty} (\tanh r)^n N_n |\Psi_n\rangle_1 |n\rangle_2$$

where normalized state $|\Psi_n\rangle$ has a form

$$|\Psi_n\rangle = A^{+n} |0\rangle / (N_n \sqrt{n!}) = (1/N_n)$$
$$\left(|n\rangle + \sum_{l=1}^{n} ((-1)^l \delta^{*l} \sqrt{n(n-1)(n-2)...(n-l+1)} / l!) |n-l\rangle\right) \tag{63}$$

and normalization factor is given by

$$N_n = \left(1 + \sum_{l=1}^{n} \frac{|\delta|^{2l} n(n-1)(n-2)...(n-l+1)}{|l|^2}\right)^{1/2}. \tag{64}$$

Now, expression (62) can be named $\alpha$ – representation of the TMSV (36). The representation is dependent on parameters $\delta$ and $r$. The displacement amplitudes $\alpha^* \tanh r$ and $\alpha$ in neighboring correlated modes of the TMSV differ from each other. We can swap the displacement amplitudes in the modes. The expression (62) is transformed to initial form of the TMSV (36) ($0$ – representation of the TMSV) in the case of $\alpha = 0$ ($\delta = 0$). The number states are the Schmidt bases for the TMSV and the number of non-zero values of the decomposition (36) is the Schmidt number [1]. Respectively, the decomposition (62) can be recognized as non-Schmidt. The parameter $\delta$ approaches to zero $\delta \to 0$ when $r \to \infty$ to result in

$$|\Psi\rangle_{12} = (1/\cosh r) D_1(\alpha^*) D_2(\alpha) \sum_{n=0}^{\infty} |n\rangle_1 |n\rangle_2. \tag{65}$$

Limit case (65) corresponds to the original (perfectly correlated and maximally entangled, but unphysical) EPR state [23] in $\alpha$ – representation.

Amplitudes of the TMSV in $\alpha$ – representation are given by

$$b_n(\delta, r) = \left(\exp(-(\sinh r)^2 |\delta|^2 / 2)/\cosh r\right)(\tanh r) N_n(\delta, r). \tag{66}$$

They are even functions relatively of sign change of the classical parameter $\alpha$

$$b_n(\delta, r) = b_n(-\delta, r). \tag{67}$$

Probability to look for correlated displaced number states $D(\alpha^* \tanh r) D_2(\alpha) |n\rangle_1 |n\rangle_2$ in TMSV is given by

$$P_n(\delta, r) = (\tanh r)^{2n} |N_n|^2 \exp\left(-(\sinh r)^2 |\delta|^2\right)/(\cosh r)^2. \tag{68}$$

It is possible to show normalization condition

$$\sum_{n=0}^{\infty} P_n(\delta, r) = 1 \tag{69}$$

is accomplished for any values of $\delta$ and $r$. Probability distributions $P_n(\delta, r)$ in dependency on $n$ for different values of $\delta$ and $r$ are shown in figure 4(a-f). The probability distribution $P_n(\delta, r)$ has maximum value for some value of $n$. Maximal value of $P_n(\delta, r)$ depends on $\delta$ and $r$ in complex fashion.

## 3. Elementary gates with superposed coherent states

The mathematical formalism developed in the previous section can become the basis for consideration of issues of generating entangled states. The idea of generating new states is based on $\alpha-$representation of the input state. The input state is shifted by a certain amount which is determined by the needs of the task. Thereafter, the measurement in the auxiliary mode is executed to generate a conditioned state which is determined by $\alpha-$representation of the initial state (Fig. 3(a)). For example, $\alpha-$representation of the TMSV (62) is used to conditionally generate state being result of action of the operator $A^+$ (4) with amplitude $\delta$ in power $n$ affecting vacuum $A^{+n}|0\rangle$. The state in unmeasured mode of a composite system (62) is generated provided that displaced number state $|n, \alpha\rangle$ is measured in the neighboring mode. In particular, it was experimentally shown in [24] that use of squeezed vacuum and photon subtraction technique combined with displacement operator generates arbitrary superposition of squeezed vacuum and squeezed single photon with high precision. Efficiency of the $\alpha-$representation is not restricted by the example.

## 3.1 Two-qubit controlled-sign gate with hybrid states

Choose a superposition of coherent states to displace the initial state before measurement in auxiliary mode. Symmetry properties of the amplitudes regarding sign change of the displacement amplitude in $\alpha-$representation can be useful for generation of new superposed states. Analyze it on example of TMSV (62). Consider in the superposition (62) the term with $n=1$ for two values of the parameter $\alpha$ equal modulo but opposite on sign

$$D_1(\alpha^* \tanh s) D_2(\alpha)(a^+ - \delta^*)|0\rangle_1 |1\rangle_2, \tag{70}$$

$$D_1(-\alpha^* \tanh s) D_2(-\alpha)(a^+ + \delta^*)|0\rangle_1 |1\rangle_2. \tag{71}$$

Deterministic operation of the displacement in neighboring modes by values $-\alpha^* \tanh s$, $-\alpha$ for the state (70) and $\alpha^* \tanh s$, $\alpha$ for the state (71) followed by measurement of single photon (probabilistic operation) in second mode is required to generate qubits

$$|\varphi_1\rangle = -a|0\rangle + b|1\rangle, \tag{72}$$

$$|\varphi_2\rangle = a|0\rangle + b|1\rangle, \tag{73}$$

where $a = \delta^*/\sqrt{1+|\delta|^2}$ and $b = 1/\sqrt{1+|\delta|^2}$. Thus, superposition of the coherent states as control qubit is used to deterministically displace TMSV being the target qubit. Initially, the control and target qubits are separable. Measurement of the single photon in auxiliary mode entangles them due to constructive interference. This idea can be the basis for the implementation of elementary quantum gates. The output state is a hybrid and it is formed by the states from different two-dimensional Hilbert spaces.

Consider the optical implementation of the idea. The scheme in Fig. 5 is a Mach-Zehnder interferometer in arms of which control qubit additionally interacts with TMSV through the beam splitters. TMSV occupies modes 3 and 4. Similar scheme with second-order crystals in arms of the interferometer was used in [25] in order to conditionally generate maximally entangled state of two photons provided that pumping photon is measured. Input qubit is in mode 1, while mode 2 is launched to the interferometer in vacuum state. Initial qubit splits on input beam splitter $B'_{12}$ and travels simultaneously along both interferometer's modes to the output beam splitter $B_{12}$ to combine on it. Depending on the beam splitter parameters, in general case, output qubit in both modes goes out from the interferometer. But if the output beam splitter is chosen so that $B_{12} = (B'_{12})^{-1}$, then output qubit certainly comes out in the input mode.

Unbalanced beam splitters (UBSs) $B'_{12}$ and $B_{12}$ are described with the following matrixes

$$B'_{12} = \begin{vmatrix} t_1 & r_1 \\ -r_1 & t_1 \end{vmatrix}, \tag{74}$$

$$B_{12} = \begin{vmatrix} t_1 & -r_1 \\ r_1 & t_1 \end{vmatrix}, \tag{75}$$

with amplitudes of transparency and reflectivity

$$t_1 = \tanh s / \sqrt{1 + (\tanh s)^2}, \tag{76}$$

$$r_1 = 1 / \sqrt{1 + (\tanh s)^2}, \tag{77}$$

respectively, where $s$ is a squeezing parameter of the TMSV. Two UBSs $B_{13}$ and $B_{24}$ are inserted inside the interferometer to organize interaction of the control coherent qubit with TMSV

$$B_{13} = \begin{vmatrix} t & -r\exp(-i\varphi) \\ r\exp(i\varphi) & t \end{vmatrix}, \tag{78}$$

$$B_{24} = \begin{vmatrix} t & -r\exp(i\varphi) \\ r\exp(-i\varphi) & t \end{vmatrix}, \tag{79}$$

where $t \to 1$ is an amplitude of transparency, $r \to 0$ is an amplitude of reflectivity and $\varphi$ is an additional phase shift.

The coherent states

$$|0_L\rangle = |0, \alpha t\sqrt{1 + (\tanh s)^2}/r\rangle, \tag{80}$$

$$|1_L\rangle = |0, -\alpha t\sqrt{1 + (\tanh s)^2}/r\rangle \tag{81}$$

are chosen as the base elements in two-dimensional Hilbert space of the control qubit $(\alpha > 0)$. The input base states are asymptotically orthogonal since their overlapping $|\langle 0_L | 1_L \rangle|^2 = \exp\left(-4\alpha t\sqrt{1 + (\tanh s)^2}/r\right) \to 0$ exponentially drops with argument growing. The condition $r \to 0$ guaranties almost perfect orthogonality of the coherent states (80,81). Therefore, we can consider the coherent states (80,81) orthogonal. Superposition of the base elements can be written as

$$|\Psi_1\rangle = a|0_L\rangle + b|1_L\rangle =$$
$$\left|\Psi_{ab}\left(\alpha t\sqrt{1+(\tanh s)^2}/r\right)\right\rangle = a\left|0,\alpha t\sqrt{1+(\tanh s)^2}/r\right\rangle + b\left|0,-\alpha t\sqrt{1+(\tanh s)^2}/r\right\rangle, \quad (82)$$

where normalization condition $|a|^2 + |b|^2 = 1$ holds. Another two-dimensional Hilbert space is formed from vacuum and single photon

$$|0_{L1}\rangle = |0\rangle, \quad (83)$$
$$|1_{L1}\rangle = |1\rangle, \quad (84)$$

being the base elements. Arbitrary superposition is defined by

$$|\Psi_2\rangle = a_1|0_{L1}\rangle + b_1|1_{L1}\rangle = a_1|0\rangle + b_1|1\rangle, \quad (85)$$

with $|a_1|^2 + |b_1|^2 = 1$.

Finally, we can write the following chain of transformations in figure 5

$$P_1(\pi)M_4^{(1)}B_{12}B_{13}B_{24}B_{12}''\left(\left|\Psi_{ab}\left(\alpha t\sqrt{1+(\tanh s)^2}/r\right)\right\rangle_1|0\rangle_2 S(r)|00\rangle_{34}\right) \rightarrow$$
$$a|0,\gamma\rangle_1(a_1|0\rangle_2 + b_1|1\rangle_2) + b|0,-\gamma\rangle_1(-a_1|0\rangle_2 + b_1|1\rangle_2) = \qquad (86)$$
$$aa_1|0,\gamma\rangle_1|0\rangle_2 + ab_1|0,\gamma\rangle_1|1\rangle_2 - ba_1|0,-\gamma\rangle_1|0\rangle_2 + bb_1|0,-\gamma\rangle_1|1\rangle_2$$

where $M_4^{(1)} = (|1\rangle\langle 1|)_4$ is a projection operator on single photon state in mode 4 and $\gamma = \alpha\sqrt{1+(\tanh s)^2}/r$ is the size of the output coherent state that is different from the size of the input base states (80,81) by a factor $t \approx 1$. Additional phase shifter on $\pi$ $P(\pi) = \exp(ia^+ a)$ in coherent mode is used to change sign of the coherent state $|0,\alpha\rangle \leftrightarrow |0,-\alpha\rangle$.

The final state (86) is the output state of controlled-sign gate provided that permutable base $|0_{L1}\rangle \leftrightarrow |1_{L1}\rangle$ for target qubit (83,84) is used. The expression (86) can be rewritten as

$$CZ(|\Psi_1\rangle|\Psi_2\rangle) = CZ(a|0_L\rangle + b|1_L\rangle)_1(a_1|0_{L1}\rangle + b_1|1_{L1}\rangle)_2 = \qquad (87)$$
$$aa_1|0_L\rangle_1|0_{L1}\rangle_2 + ab_1|0_L\rangle_1|1_{L1}\rangle_2 + ba_1|1_L\rangle_1|0_{L1}\rangle_2 - bb_1|1_L\rangle_1|1_{L1}\rangle_2,$$

or in matrix representation

$$CZ\begin{vmatrix} aa_1 \\ ab_1 \\ ba_1 \\ bb_1 \end{vmatrix} = \begin{Vmatrix} 1 & 0 & 0 & 0 \\ 0 & 1 & 0 & 0 \\ 0 & 0 & 1 & 0 \\ 0 & 0 & 0 & -1 \end{Vmatrix} \begin{vmatrix} aa_1 \\ ab_1 \\ ba_1 \\ bb_1 \end{vmatrix} = \begin{vmatrix} aa_1 \\ ab_1 \\ ba_1 \\ -bb_1 \end{vmatrix}, \qquad (88)$$

where the control qubit $|\Psi_1\rangle$ can be considered as macroscopic and the target qubit $|\Psi_2\rangle$ is microscopic.

Given operation can be directly realized with the target qubit (85) additionally entangled with another auxiliary qubit (mode 3)

$$|\Psi_2\rangle = a_1|01\rangle_{23} + b_1|10\rangle_{23}. \qquad (89)$$

The control coherent qubit (82) interacts with auxiliary mode on UBS with the transmittance $T \rightarrow 1$ (Fig. 3(b)) to deterministically displace the target state by equal modulo but opposite by sign values. Next registration of the single photon in the auxiliary mode generates entangled state being outcome of the controlled-sign gate

$$M_1^{(1)}B_{13}(|\Psi_1\rangle|\Psi_2\rangle)(|\Psi_{ab}(\alpha t/r)\rangle_1(a_1|01\rangle_{23} + b_1|10\rangle_{23})) \rightarrow$$
$$aa_1|0,\alpha t/r\rangle_1|0\rangle_2 + ab_1|0,\alpha t/r\rangle_1|1\rangle_2 + ba_1|0,-\alpha t/r\rangle_1|0\rangle_2 - bb_1|0,-\alpha t/r\rangle_1|1\rangle_2, \qquad (90)$$

provided that $\alpha = (\sqrt{5} - 1)/2$. Conclusion of the value for displacement amplitude is based on the transformation matrix (13), namely, on its zeroth and first rows in the case of $\alpha' = 0$. The next step is to find a value of $\alpha$ at which the matrix elements $c_{01}$ and $c_{11}$ would be equal to each other in absolute value. Analysis of the matrix elements allows for one to find the appropriate value of $\alpha$. Since such value of $\alpha$ exists it allows to deterministic shift of the state in the third mode by $\pm \alpha$ with the subsequent registration of a single photon to obtain final state (90).

It is known that deterministic single qubit gates with photons can be readily performed with simple linear optics but the deterministic two-qubit gate requires the introduction of a very large nonlinearity. Implementation of the two-qubit gates with photons is hardly possible due to the fact that photons do not interact with each other. As an alternative to the deterministic scheme, one may generate the transformation probabilistically using a measurement-induced nonlinearity by applying projectors like projection operator on single photon used in figure 5. Solution of the problem with maximum efficiency and minimum number of resources is possible with help of the use of optical qubits having a certain size. In addition, the proposed optical circuit implements a two-qubit-operation without the use of additional quantum state (channel).

**3.2 Hadamard matrix with superposed coherent states**

Concerning the SCS, it is difficult to control the states at the local level. Consider alternative possibility to manipulate the SCSs. The same optical scheme is also used as in the case of the two-qubit transformation in figure 5. Consider the partial case of $a = b = 1/\sqrt{2}$ ($a = -b = 1/\sqrt{2}$), $\delta = \pm 1$ and as consequence $a_1 = b_1 = 1/\sqrt{2}$ in equation (86). Then, the balanced hybrid states are generated

$$|\Phi_+\rangle_{12} = \frac{1}{2}\begin{pmatrix} \left|0, \alpha\sqrt{1+(\tanh s)^2}/r\right\rangle_1 (|0\rangle + |1\rangle)_2 + \\ \left|0, -\alpha\sqrt{1+(\tanh s)^2}/r\right\rangle_1 (-|0\rangle + |1\rangle)_2 \end{pmatrix}, \quad (91)$$

$$|\Phi_-\rangle_{12} = \frac{1}{2}\begin{pmatrix} \left|0, \alpha\sqrt{1+(\tanh s)^2}/r\right\rangle_1 (|0\rangle + |1\rangle)_2 - \\ \left|0, -\alpha\sqrt{1+(\tanh s)^2}/r\right\rangle_1 (-|0\rangle + |1\rangle)_2 \end{pmatrix}. \quad (92)$$

The two-mode states are entangled and orthogonal to each other
$$\langle \Phi_- | \Phi_+ \rangle = 0, \quad (93)$$
and can be considered as the base elements of the output Hilbert space
$$|0_{L2}\rangle = |\Phi_+\rangle_{12}, \quad (94)$$
$$|1_{L2}\rangle = |\Phi_-\rangle_{12}. \quad (95)$$

The values of the parameter $\delta = \pm 1$ correspond to the displacement amplitudes (61)
$$\alpha = \pm \sinh s \cosh s. \quad (96)$$

Using the expression (86), one obtains

$$P_1(\pi)M_4^{(1)}B_{12}B_{13}B_{24}B_{12}''\left(\left|\Psi_{ab}\left(\alpha t\sqrt{1+(\tanh s)^2}/r\right)\right\rangle_1|0\rangle_2 S(r)|00\rangle_{34}\right) \to \qquad (97)$$
$$((a+b)/\sqrt{2})|\Phi_+\rangle_{12} + (a-b)|\Phi_-\rangle_{12}/\sqrt{2}$$

provided that detector in mode 4 registered a click. The outcome can be fully recognized as outcome of the Hadamard matrix

$$H = \frac{1}{\sqrt{2}}\begin{vmatrix} 1 & 1 \\ 1 & -1 \end{vmatrix} \qquad (98)$$

acting on the state

$$H|\Psi_1\rangle = H(a|0_L\rangle + b|1_L\rangle) = ((a+b)/\sqrt{2})|0_{L2}\rangle + ((a-b)/\sqrt{2})|1_{L2}\rangle, \qquad (99)$$

or the same on column vector of the input amplitudes

$$H\begin{vmatrix} a \\ b \end{vmatrix} = \frac{1}{\sqrt{2}}\begin{vmatrix} a+b \\ a-b \end{vmatrix}, \qquad (100)$$

where input and output base elements are chosen from different two-dimensional Hilbert spaces.

Reverse operation can be also realized. To show it one should make use of $\alpha$-representation of superposition of vacuum and single photon (29,30). Mixing of the superposition with superposed coherent states followed by measurement of single photon gives

$$H^{-1}(a|0_{L2}\rangle + b|1_{L2}\rangle) = H^{-1}(a|\Phi_+\rangle_{12} + b|\Phi_-\rangle_{12}) \to (a+b/\sqrt{2})|0_L\rangle + (a-b/\sqrt{2})|1_L\rangle =$$
$$((a+b)/\sqrt{2})|0,\alpha t\sqrt{1+(\tanh s)^2}/(Tr)\rangle_1 + ((a-b)/\sqrt{2})|0,-\alpha t\sqrt{1+(\tanh s)^2}/(Tr)\rangle \qquad (101)$$

where $H^{-1}$ is a matrix reverse to Hadamard $H = H^{-1}$ and $T = t^2$. Output state is an outcome of Hadamard matrix transforming hybrid state to the superposed coherent states.

Consider implementation of the Hadamard matrix between vector spaces of equal dimension, namely, between one-mode macroscopic and microscopic vector spaces. The base coherent states (80,81) can be considered as macroscopic due to big values of their sizes. The basic elements of the second vector space are the microscopic states (83,84). The same optical scheme in figure 5 is used with additional projection operator $M_1^{(n)} = |n\rangle\langle n|$ in first mode. The chain of the transformations becomes

$$M_1^{(n)}M_4^{(1)}B_{12}B_{13}B_{24}B_{12}''\left(\left|\Psi_{ab}\left(\alpha t\sqrt{1+(\tanh s)^2}/r\right)\right\rangle_1|0\rangle_2 S(r)|00\rangle_{34}\right) \to \qquad (102)$$
$$a(|0\rangle+|1\rangle)_1/\sqrt{2} + (-1)^n b(-|0\rangle+|1\rangle)_1/\sqrt{2} = ((a+(-1)^{n+1}b)/\sqrt{2})|0\rangle_1 + ((a+(-1)^n b)/\sqrt{2})|1\rangle_1$$

Output state depends on number (even or odd) of the measured photons in first mode. If odd number of photons is measured, then output is exactly outcome of the Hadamard matrix (100). If even number of photons is measured, then the final state is output of the matrix $U_y(Q = -\pi/2)$ that belongs to class of unitary matrices originating from Pauli matrixes when they are exponentiated [1].

Reverse action may be realized with help of the single-mode squeezing operator [22]

$$S_1(r) = \exp(r(a_1^{+2} - a_1^2)/2), \qquad (103)$$

where $r$ is a squeezing parameter. Squeezed vacuum and squeezed single photon may approximate even and odd SCSs of a certain size [10, 11, 17, 18]

$$S(r)|0\rangle \to N_+(\alpha)(|0,\alpha\rangle + |0,-\alpha\rangle), \qquad (104)$$
$$S(r)|1\rangle \to N_-(\alpha)(|0,\alpha\rangle - |0,-\alpha\rangle). \qquad (105)$$

Fidelity of the generated states drops with increase of amplitude $\alpha$. Increase of the SCSs size is achieved by application of photon subtraction technique to initial Gaussian states [17, 18]. We can only conjecture that squeezed superpositions of vacuum and single photon may approximate coherent states superpositions by analogy with (104,105)

$$S(r)(a|0\rangle + b|1\rangle)/\sqrt{2} \to ((a+b)/\sqrt{2})|0,\alpha\rangle + ((a-b)/\sqrt{2})|0,-\alpha\rangle.$$

## 4. Realization of the elementary gates in realistic scenario

We have considered implementation of elementary gates in conjecture of ideal projective measurement. As APD cannot distinguish number of incoming photons it induces additional contribution of higher-order terms of TMSV (62) to the output state. It leads to decrease of unity fidelity of the generated states. Single photon must be registered by APD in mode 4 for implementation of the gates. Therefore, small values of the squeezing parameter $s \ll 1$ must be taken to ensure dominance of the states $|0\rangle$ and $|1\rangle$ over other states in distribution of the TMSV. Contribution of the higher order number states becomes negligible in the case of $s \ll 1$. When APD registers click in mode 4, it implies with almost unity probability that single photon was measured in the case of $s \ll 1$. For example, figure 6 shows plots of relations of single photon probability to probabilities of higher order terms of the TMSV $P_1(\delta=1,s)/P_k(\delta=1,s)$ with $k$ from 2 up to 7 in dependency on squeezing parameter $s$ for the case of $\delta = 1$.

Consider realization of the projective measurement in mode 1 with APD. Coherent states with large amplitudes $\pm \alpha \sqrt{1 + \tanh^2 s}/r$ go towards APD in mode 1. Therefore, one needs to decrease their amplitudes before measurement in the mode. Suppose the absorbing medium transforms the coherent states as $|0,\pm\alpha\rangle \to |0,\pm\alpha/A\rangle$, where $A$ is an absorbing coefficient. The parameter can be chosen in such a way the absolute amplitude of the coherent states to become close to zero. The coherent states may be approximated by superposition $|0,\pm\alpha\rangle \approx |0\rangle \pm \alpha|1\rangle$. Then, if APD registers some event, it means registration of single photon with large probability.

## 5. Conclusion

The development of quantum information processing has traditionally followed two separate lines of study depending on which degree of freedom or observable is used for describing participating states. One of the lines has dealt with the processing of information by the states with the discrete degree of freedom (eigenvalues of Hamiltonian are discretized). The other has been devoted to implementations based on Gaussian states such as coherent and squeezed states when we talk about continuous-variable states. Although, we can think that this separation is artificial and is caused by the experimental difficulties in interconnecting the standard technologies of two lines. It is logical to think that the displaced number states can become the bridge that will connect and combine these two lines of quantum information processing by optical methods. Indeed, apart from the fact that these states are discrete, they include a definition that assigns to the quantum states a number, which we define as their classical size (or simply size).

Additional classical degree of freedom of the displaced states gives more possibilities to manipulate the quantum states. Developed approach allows for the consideration of the mechanism of interaction between the qubit of distinct nature. The alternative mechanism of subtraction of displaced single photon gives a possibility to effectively achieve nonlinear effect on qubits comparable with that of nonlinear media. Deterministic operation of displacement of the state allows for one to extract new information from the initial state by following subtraction of single photon. As a result, output state becomes entangled and hybrid. Photon subtraction is achieved with zeroth amplitude of the displacement. Therefore, it can be considered as partial case of extraction of displaced photon. Nonzero amplitude of the extracted displaced photon enables to get information in different coding.

We considered a possibility of implementation of the elementary quantum gates with different base states based on their displaced properties. Output entangled hybrid state of the controlled-sign gate is composed of superposed coherent states (macroscopic state) and superposition of vacuum and single photon (microscopic). Hadamard transformation is realized between different two-dimensional Hilbert spaces. Input SCS can be transformed to either hybrid or two-level superposition. The method requires minimal number of resources, it works without Bell-state measurement and initial quantum channel. We have shown one APD can be used in the scheme. The gates are probabilistic and success probability of the operation can be increased.

This approach is based on a new mathematical apparatus that is developed in this paper. The transformation matrix (13) is a key moment of the mathematical method. The matrix lies in core of introduction of the $\alpha-$ representation of the arbitrary state. The representation is obtained by multiplication of the transposed transformation matrix on column vector of state amplitudes and presents decomposition of the state in terms of the displaced number states with displacement amplitude $\alpha$. Analytical expressions of the $\alpha-$ representation of superposition of vacuum and single photon TMSW are derived. Conclusion of the $\alpha-$ representation of TMSW is not trivial and requires additional mathematical transformations with matrix elements. Symmetry properties of the matrix elements, in particular TMSV, in regard to sign change of the displacement amplitude enable to manipulate the target states in such a way to generate entangled hybrid states being outcome of the controlled sign gate. So, qubits can be governed by classical information conveyed by the coherent states. The developed mathematical approach can be common in the analysis of the problems with hybrid states.

**LIST OF FIGURES**

**Figure 1**
Phase-space plot in terms of $X$ and $P$ (position and momentum of a harmonic oscillator) shows the uncertainties of Fock state $|n\rangle$ and its displaced analogue $|n,\alpha\rangle$. The uncertainties are represented by the circles of the same radius distant from each other by $|\alpha|$. The radius depends on number of the state $R=(2n+1)/4$.

**Figure 2(a-f)**
Probability distribution of the single photon state over displaced number states with different values of the displacement amplitude.

**Figure 3(a,b)**
Schematic image of realization of action of the projective operator (26) on arbitrary state $|\Psi\rangle$. The displacement operator $D(-\alpha)$ with amplitude $-\alpha$ is applied to the state before its measurement (a). The displacement operator (b) is deterministically accomplished with help of unbalanced beam splitter (UBS) with transmittance $T \to 1$.

**Figure 4(a-f)**
Probability distribution $P_{n-}(\alpha)$ of superposition of vacuum and single photon over displaced number states for different values of the displacement amplitude $\alpha$.

**Figure 5**
The optical scheme is used for realization of one- and two-qubit operations. Control coherent qubit is launched into interferometer in mode $1$, splits onto two modes in UBS $B'_{12}$, again combines on UBS $B_{12}$ and interacts with TMSV in the interim between the UBSs. TMSV occupies modes $3$ and $4$. Target qubit is initially contained in the TMSV. Interaction of the control qubit and TMSV occurs simultaneously on two UBSs $B_{13}$ and $B_{24}$. Registration of the single photon in mode $4$ guaranties generation of hybrid state.

**Figure 6**
Dependencies of relations of probability of displaced single photon to the probabilities of higher-order terms of TMSV $P_1(\delta=1,s)/P_{k+1}(\delta=1,s)$ (curve $k$), where $k$ is varied from $1$ up to $6$, on $s$.

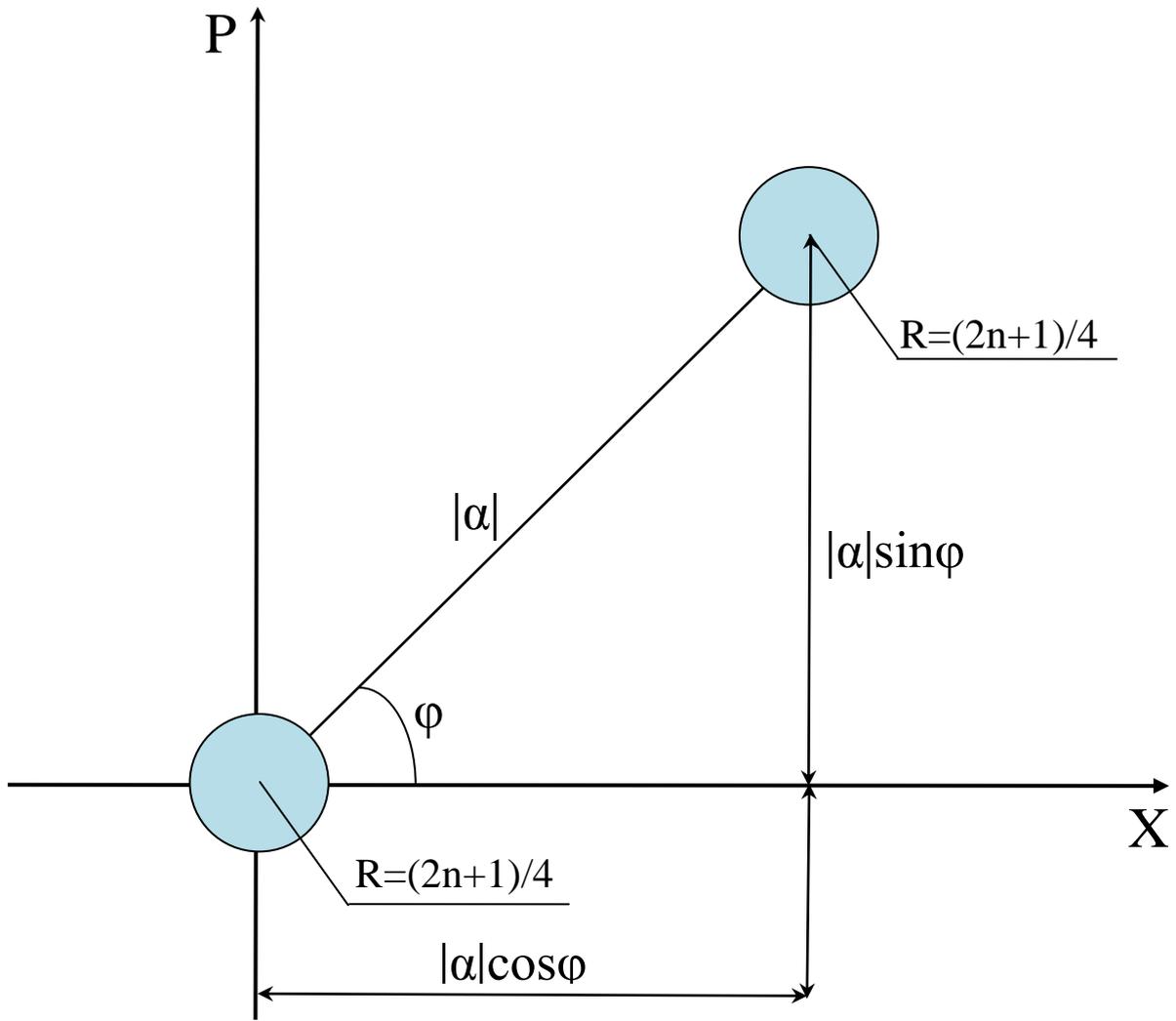

**Figure 1**

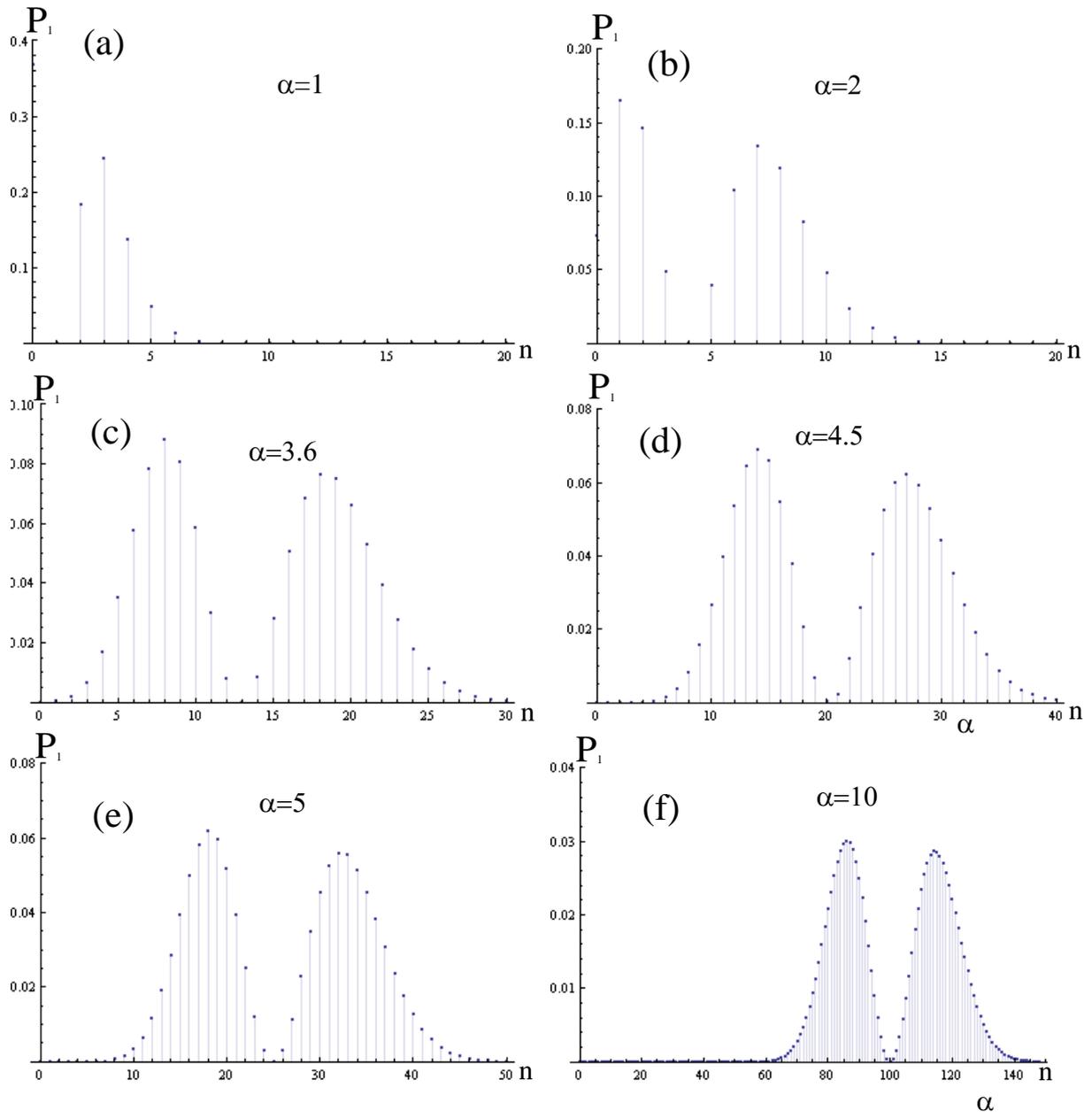

**Figure 2(a-f)**

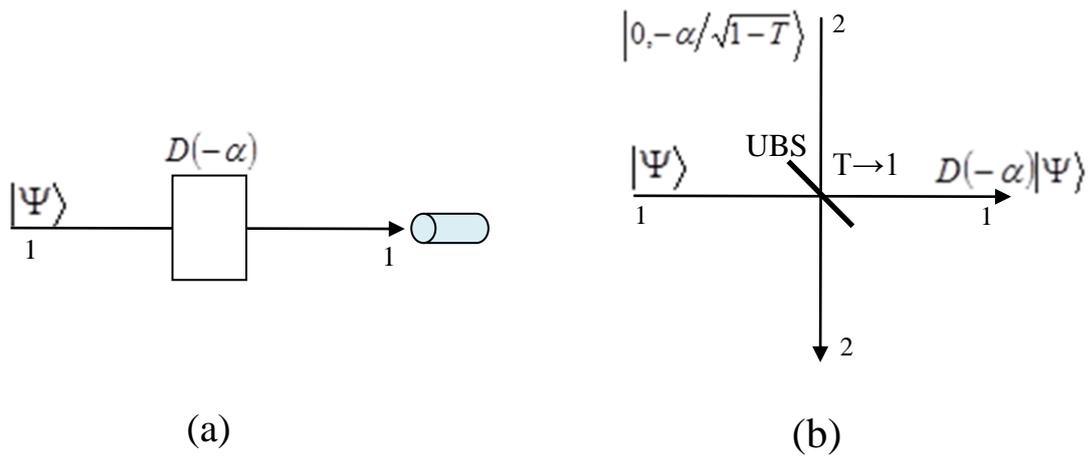

(a)    (b)

**Figure 3(a, b)**

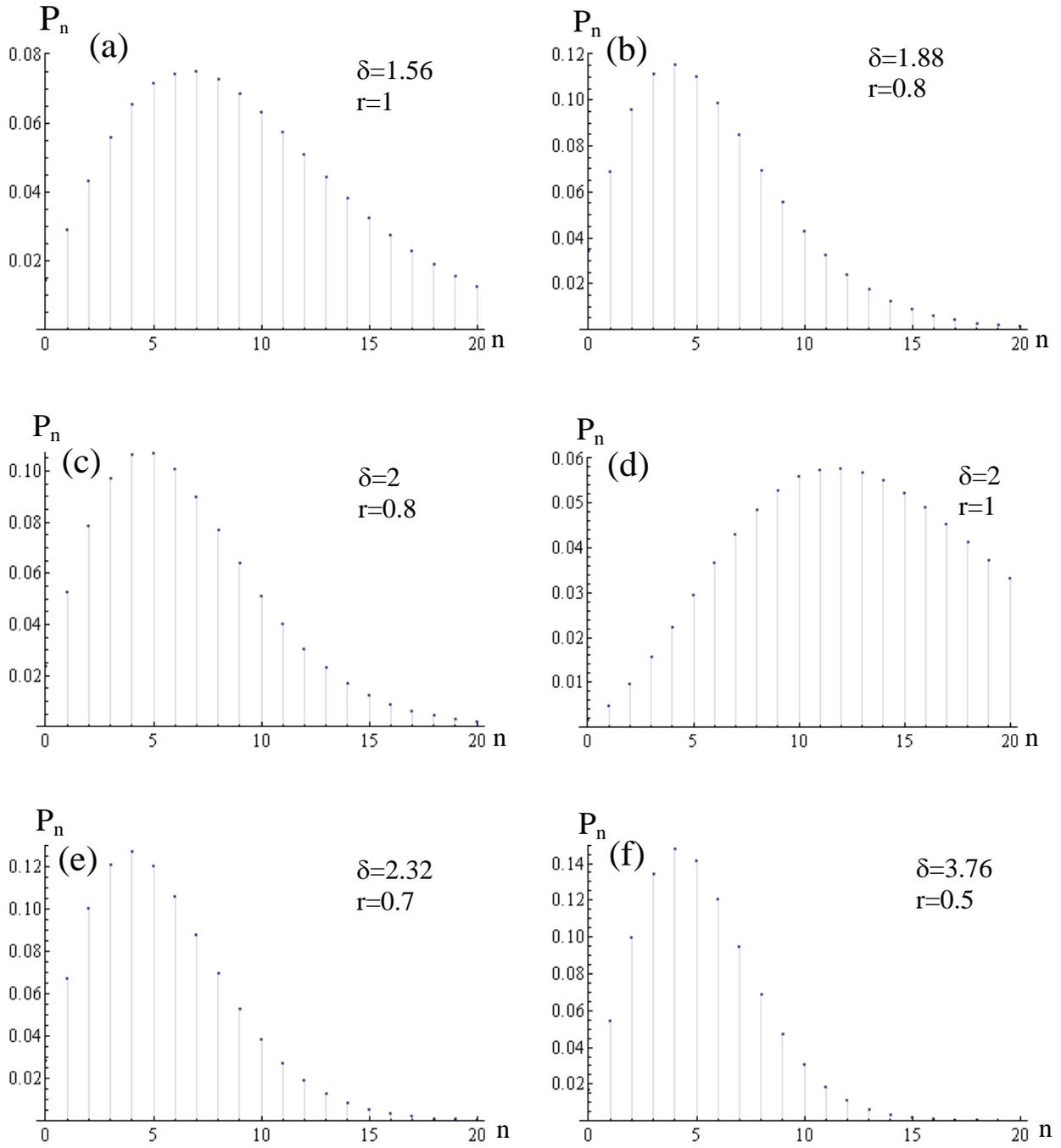

**Figure 4(a-f)**

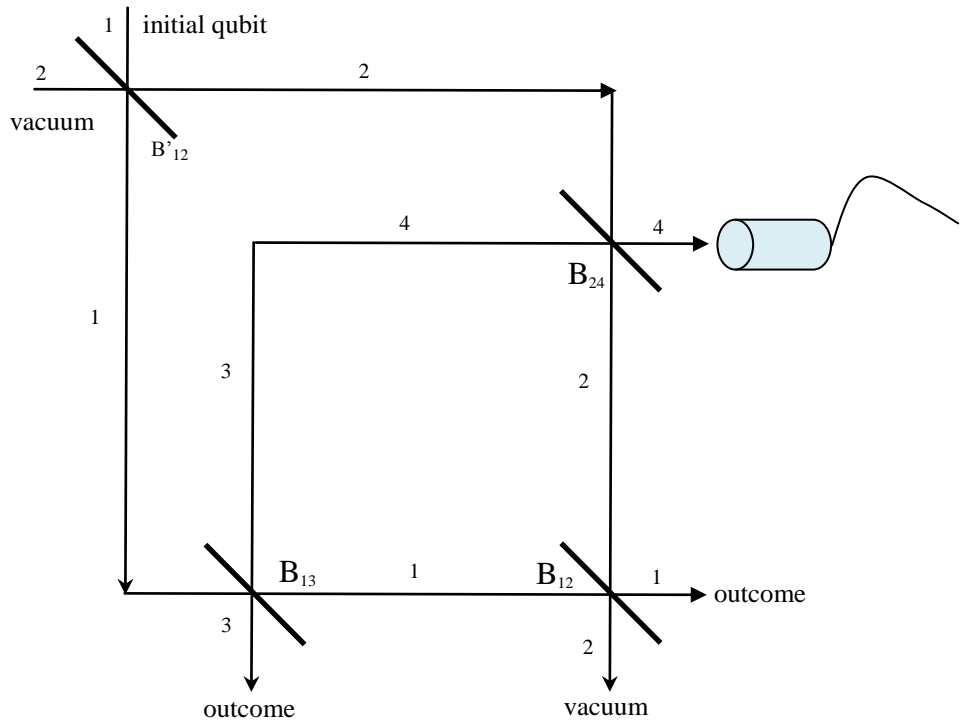

**Figure 5**

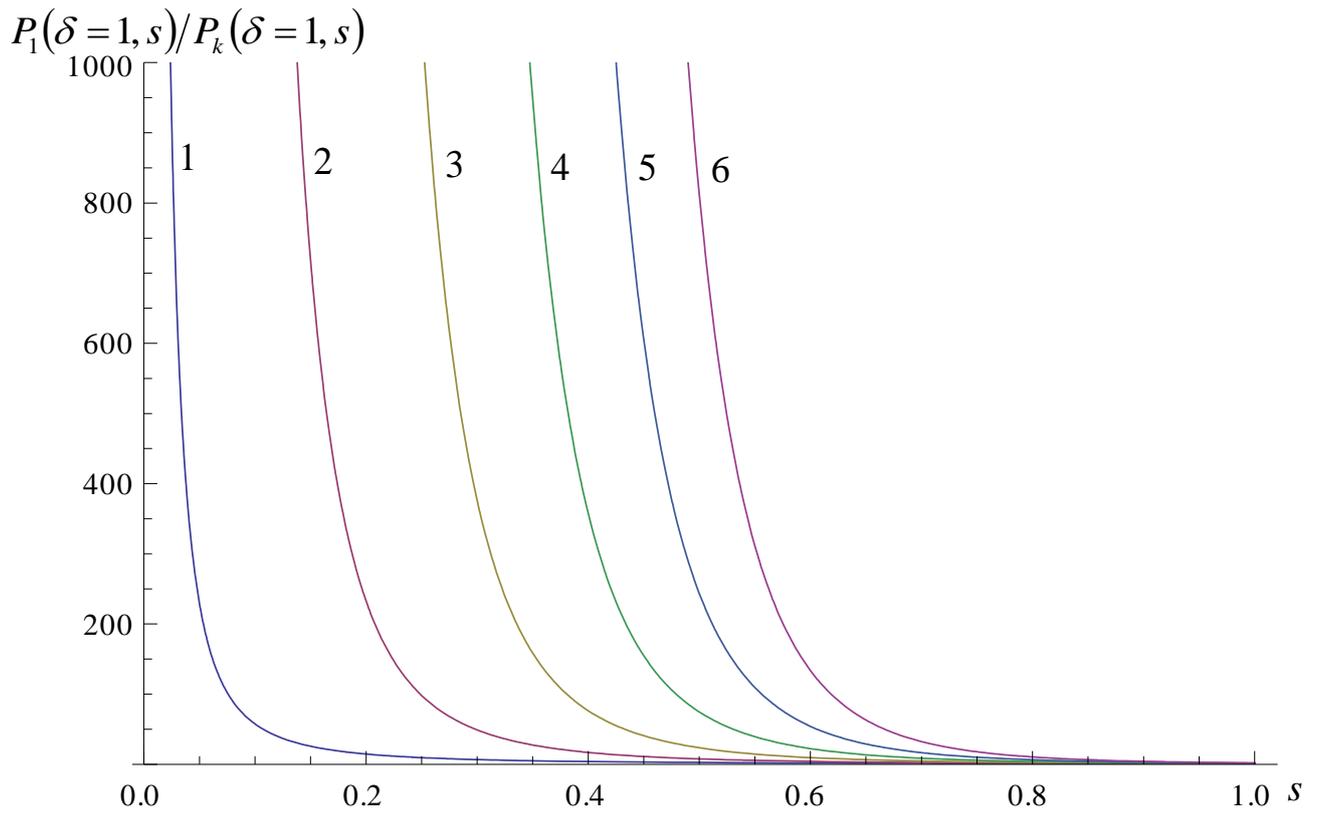

**Figure 6**